\documentclass[sigconf]{acmart} 

\AtBeginDocument{%
  }

\usepackage{multirow}
\usepackage{caption}
\usepackage{subcaption}
\usepackage{xspace}

\newcommand{\system}{{PLAID}\xspace}
\newcommand{\systemlong}{\system{} ColBERTv2\xspace}
\newcommand{\vanilla}{vanilla ColBERTv2\xspace}
\newcommand{\Vanilla}{Vanilla ColBERTv2\xspace}
\newcommand{\secref}[1]{\S\ref{#1}\xspace}

\begin{document}

\title{\system{}: An Efficient Engine for Late Interaction Retrieval}

\author{Keshav Santhanam}
\authornote{Equal contribution.}
\email{keshav2@stanford.edu}

\affiliation{%
  \institution{Stanford University}
\country{United States}
}

\author{Omar Khattab}
\authornotemark[1]
\email{okhattab@stanford.edu}
\affiliation{%
\institution{Stanford University}
\country{United States}
  }

\author{Christopher Potts}
\affiliation{%
\institution{Stanford University}
 \country{United States}
  }

\author{Matei Zaharia}
\affiliation{%
\institution{Stanford University}
\country{United States}
}

\renewcommand{\shortauthors}{Keshav Santhanam*, Omar Khattab*, Christopher Potts, and Matei Zaharia}

\begin{abstract}

Pre-trained language models are increasingly important components across multiple information retrieval (IR) paradigms. Late interaction, introduced with the ColBERT model and recently refined in ColBERTv2, is a popular paradigm that holds state-of-the-art status across many benchmarks. To dramatically speed up the search latency of late interaction, we introduce the Performance-optimized Late Interaction Driver (\system{}). Without impacting quality, PLAID swiftly eliminates low-scoring passages using a novel centroid interaction mechanism that treats every passage as a lightweight bag of centroids. \system{} uses centroid interaction as well as centroid pruning, a mechanism for sparsifying the bag of centroids, within a highly-optimized engine to reduce late interaction search latency by up to 7$\times$ on a GPU and 45$\times$ on a CPU against \vanilla{}, while continuing to deliver state-of-the-art retrieval quality. This allows the \system{} engine with ColBERTv2 to achieve latency of tens of milliseconds on a GPU and tens or just few hundreds of milliseconds on a CPU at large scale, even at the largest scales we evaluate with 140M passages.

\end{abstract}

\maketitle

\section{Introduction} \label{section:introduction}

\begin{figure}[]
\centering
\includegraphics[width=0.95\columnwidth]{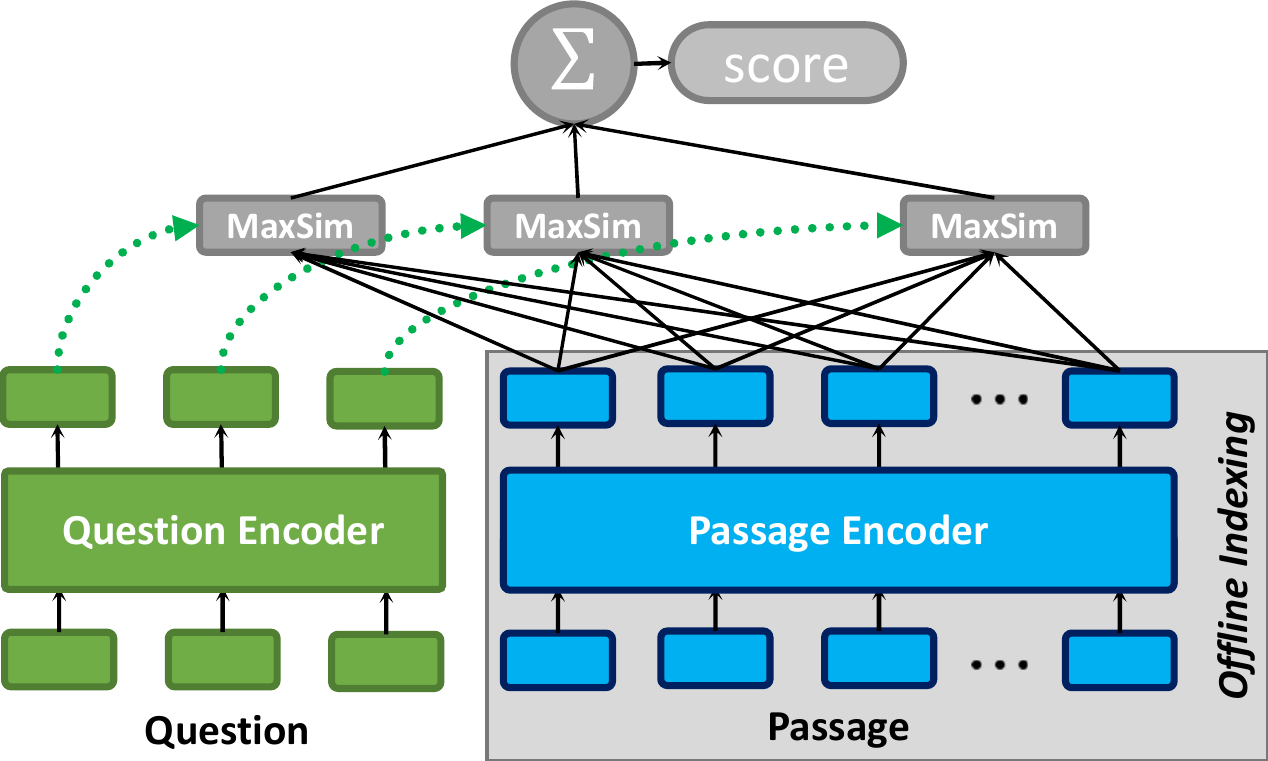}
\caption{The late interaction architecture, given a query and a passage. Diagram from \citet{khattab2021relevance} with permission.}
\label{fig:ColBERT}
\end{figure}

Recent advances in neural information retrieval (IR) have led to notable gains on retrieval benchmarks and retrieval-based NLP tasks. Late interaction, introduced in ColBERT~\cite{khattab2020colbert}, is a paradigm that delivers state-of-the-art quality in many of these settings, including passage ranking~\cite{santhanam2021colbertv2,hofstatter2022introducing,wang2021pseudo}, open-domain question answering~\cite{khattab2021relevance,li2021learning}, conversational tasks~\cite{paranjape2021hindsight,minas2022zero}, and beyond~\cite{khattab2021baleen,zhong2022evaluating}. ColBERT and its variants encode queries and documents into \textit{token-level} vectors and conduct scoring via scalable yet \textit{fine-grained} interactions at the level of tokens (Figure~\ref{fig:ColBERT}), alleviating the dot-product bottleneck of single-vector representations. The recent ColBERTv2~\cite{santhanam2021colbertv2} model demonstrates that late interaction models often considerably outperform recent single-vector and sparse representations within and outside the training domain, a finding echoed in several recent studies~\cite{lupart2022toward,zhan2022evaluating,thakur2021beir,macdonald2021single,zeng2022curriculum,thai2022relic}.

Despite its strong retrieval quality, late interaction requires special infrastructure~\cite{khattab2020colbert,lin2022proposed} for low-latency retrieval as it encodes each query and each document as a full \textit{matrix}. Most IR models represent documents as a single vector, either sparse (e.g., BM25~\cite{robertson1995okapi}; SPLADE~\cite{formal2021splade}) or dense (e.g., DPR~\cite{karpukhin2020dense}; ANCE~\cite{xiong2020approximate}), and thus mature sparse retrieval strategies like WAND~\cite{broder2003efficient} or dense kNN methods like HNSW~\cite{malkov2018efficient} cannot be applied directly or optimally to late interaction. While recent work~\cite{macdonald2021approximate,tonellotto2021query,santhanam2021colbertv2} has explored optimizing individual components of ColBERT's pipeline, an end-to-end optimized engine has never been studied to our knowledge.

We study how to optimize late-interaction search latency at a large scale, taking all steps of retrieval into account. We build on the state-of-the-art ColBERTv2 model. Besides improving quality with denoised supervision, ColBERTv2 aggressively compresses the storage footprint of late interaction. It reduces the index size by up to an order of magnitude using \textit{residual representations} (\secref{subsection:analysis:review}). In those, each vector in a passage is encoded using the ID of its nearest \textit{centroid} that approximates its token semantics---among tens or hundreds of thousands of centroids obtained through $k$-means clustering---and a \textit{quantized} residual vector.

We introduce the \textbf{P}erformance-optimized \textbf{La}te \textbf{I}nteraction \textbf{D}river (\system{}),\footnote{Code maintained at \url{https://github.com/stanford-futuredata/ColBERT}. As of May'22, \system{} lies under the branch \href{https://github.com/stanford-futuredata/ColBERT/tree/fast_search}{\texttt{fast\_search}} but will soon be merged upstream.} an efficient retrieval engine that reduces late interaction search latency by 2.5--7$\times$ on GPU and 9--45$\times$ on CPU against \vanilla{} while retaining high quality. This allows the \system{} implementation of ColBERTv2, \systemlong{}, to achieve CPU-only latency of tens or just few hundreds of milliseconds and GPU latency of few tens of milliseconds at very large scale, even on 140M passages. Crucially, \systemlong{} does so while continuing to deliver state-of-the-art retrieval quality.

To dramatically speed up search, \system{} leverages the centroid component of the ColBERTv2 representations, which is a compact integer ID per token. Instead of exhaustively scoring all passages found with nearest-neighbor search, \system{} uses the centroids to identify high-scoring passages and eliminate weaker candidates without loading their larger residuals. We conduct this in a multi-stage pipeline and introduce \textit{centroid interaction}, a scoring mechanism that treats every passage as a lightweight \textit{bag of centroid IDs}. We show that this centroid-only multi-vector search exhibits high recall without using the vector residuals (\secref{subsection:analysis:centroids}), allowing us to reserve full scoring to a very small number of candidate passages. Because the centroids come from a fixed set (i.e., constitute a discrete vocabulary), the distance between the query vectors and all centroids can be computed once during search and \textit{re-used} across all bag-of-centroids passage representations. This allows us to further leverage the centroid scores for \textit{centroid pruning}, which sparsifies the bag of centroid representations in the earlier stages of retrieval by skipping centroid IDs that are distant from all query vectors.

In the \system{} engine, we implement centroid interaction and centroid pruning and implement optimized yet modular kernels for the data movement, decompression, and scoring components of late interaction with the residual representations of ColBERTv2 (\secref{section:implementation}). We extensively evaluate the quality and efficiency of \system{} within and outside the training domain (on MS MARCO v1~\cite{nguyen2016ms} and v2~\cite{craswell2022overview}, Wikipedia, and LoTTE~\cite{santhanam2021colbertv2}) and across a wide range of corpus sizes (2M--140M passages), search depths ($k$=10, 100, 1000), and hardware settings with single- and multi-threaded CPU and with a GPU (\secref{subsection:eval:end_to_end}). We also conduct a detailed ablation study to understand the empirical sources of gains among centroid interaction, centroid pruning, and our faster kernels (\secref{subsection:eval:ablation}).

In summary, we make the following contributions:

\begin{enumerate}
    \item We analyze centroid-only retrieval with ColBERTv2, showing that a pruned bag-of-centroids representation can support high-recall candidate generation (\secref{section:analysis}).
    
    \item We propose \system{}, a retrieval engine that introduces centroid interaction and centroid pruning as well as optimized implementations of these techniques for dramatically improving the latency of late-interaction search (\secref{section:centroid_scoring_pipeline}).

    \item We extensively evaluate \system{} and conduct a large-scale evaluation up to 140M passages, the largest to our knowledge with late-interaction retrievers (\secref{section:evaluation}).
\end{enumerate}

\section{Related Work} \label{section:background}

\subsection{Neural IR} \label{subsection:background:late_interaction}

The IR community has introduced many neural IR models based on pre-trained Transformers. Whereas early models were primarily cross-encoders~\cite{nogueira2019passage,macavaney:sigir2019-cedr} that attend jointly to queries and passages, many subsequent models target higher efficiency by producing independent representations for queries and passages. Some of those produce sparse term weights~\cite{dai2020context,mallia2021learning}, whereas others encode each passage or query into a single vector ~\cite{xiong2020approximate,karpukhin2020dense,qu2021rocketqa} or multi-vector representation (the class we study; ~\cite{khattab2020colbert,santhanam2021colbertv2,khattab2021relevance,humeau2020polyencoders,gao2021coil}). These choices make different tradeoffs about efficiency and quality: whereas sparse term weights and single-vector models can be particularly lightweight in some settings, multi-vector late interaction~\cite{khattab2020colbert} can often result in considerably stronger quality and robustness. Orthogonal to the choice of modeling query--document interactions, researchers have improved the supervision for neural models with harder negatives~\cite{khattab2021relevance,xiong2020approximate,zhan2020learning} as well as distillation and denoising~\cite{qu2021rocketqa,ren2021rocketqav2,hofstatter2020improving}, among other approaches. Our work extends ColBERTv2~\cite{santhanam2021colbertv2}, which combines late interaction modeling with hard negative and denoising supervision to achieve state-of-the-art quality among standalone retrievers.

\begin{figure}
    \centering
     \includegraphics[scale=0.35]{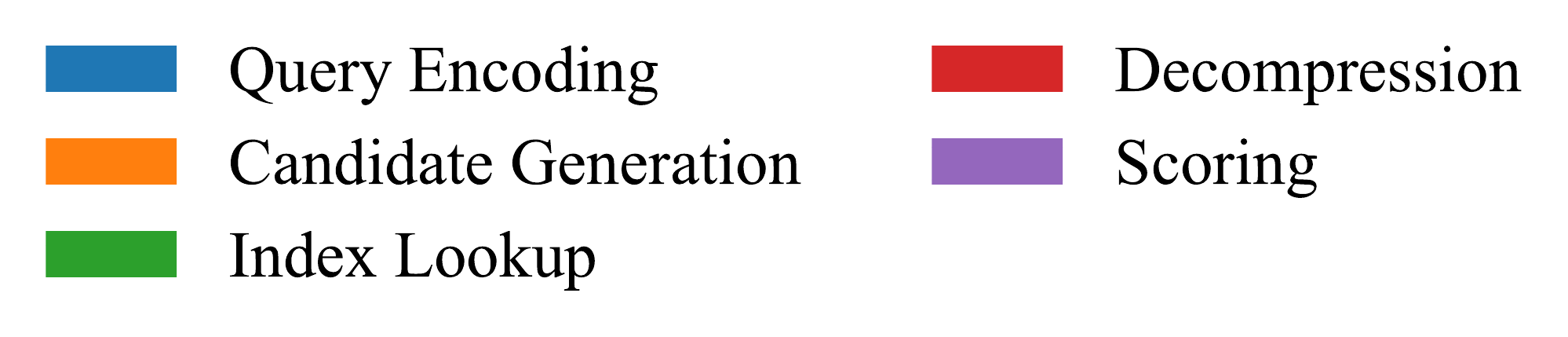} \\
    \begin{subfigure}[b]{\columnwidth}
    \centering
    \includegraphics[scale=0.35]{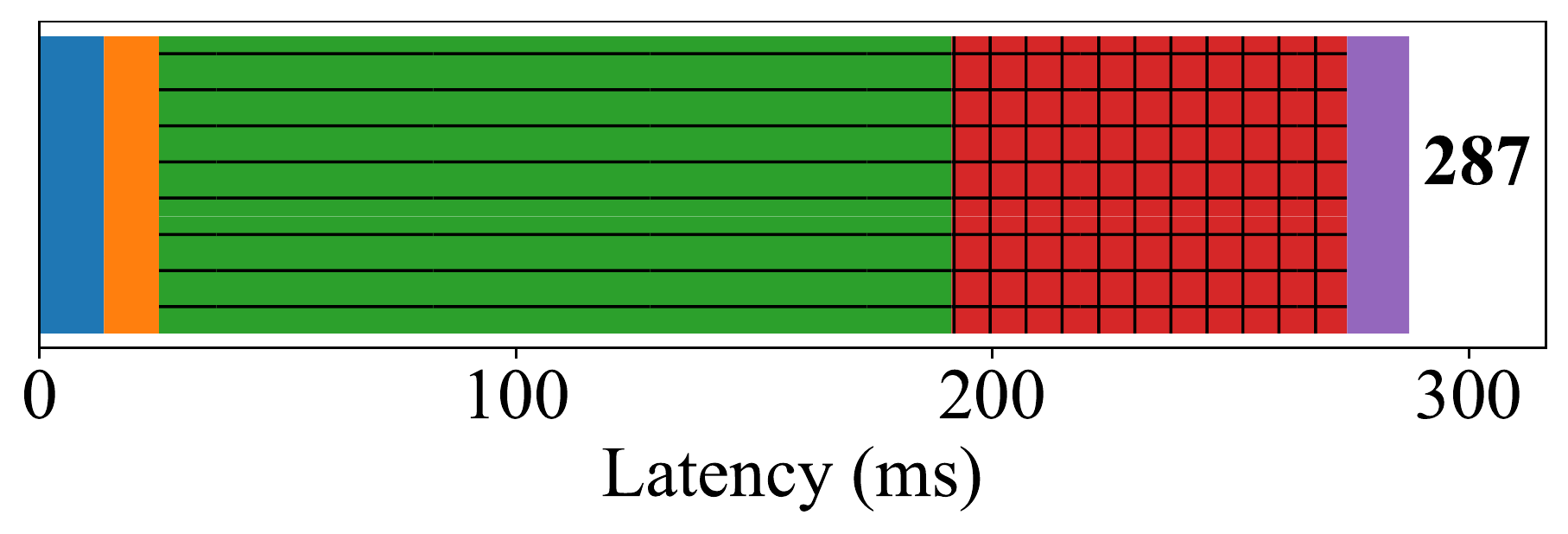}
    \caption{Vanilla ColBERTv2 (\texttt{nprobe}=4, \texttt{ncandidates}=$2^{16}$).}
    \label{fig:msmarco_baseline_latency_breakdown}
    \end{subfigure}
     \par\bigskip
      \par\bigskip
    \begin{subfigure}[b]{\columnwidth}
    \centering
    \includegraphics[scale=0.35]{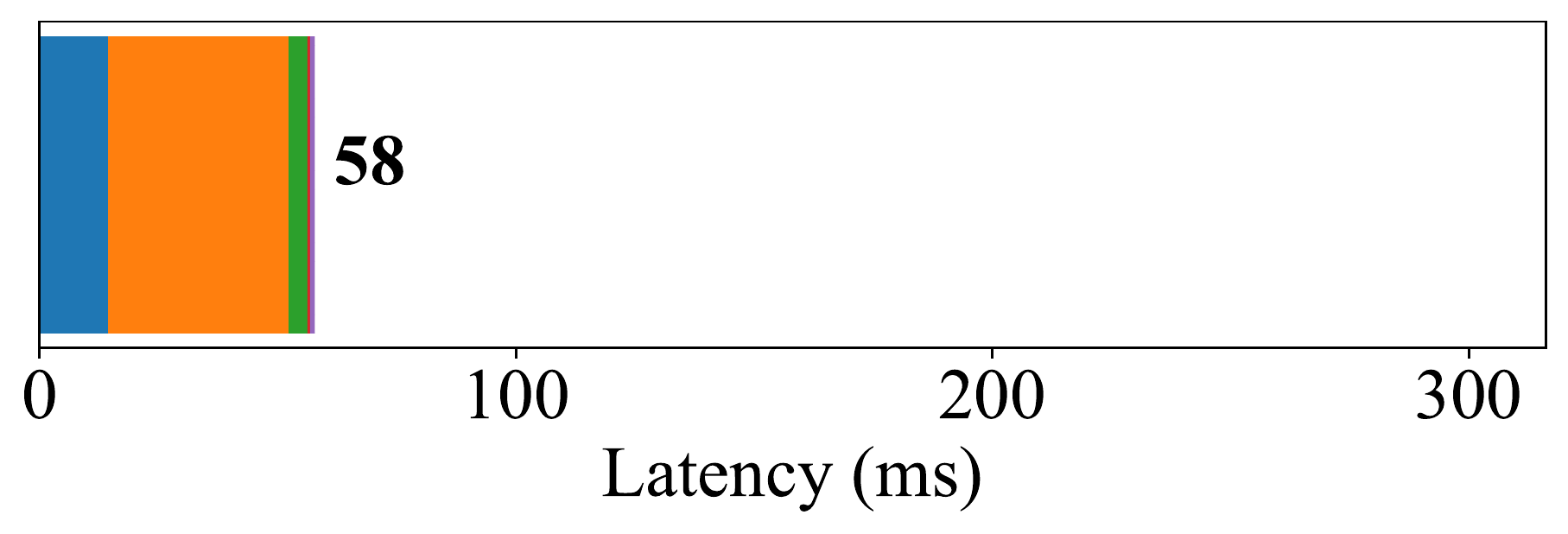}
    \caption{\systemlong{} ($k=1000$)}
    \label{fig:msmarco_plaid_latency_breakdown}
    \end{subfigure}
    \caption{Latency breakdown of MS MARCO v1 dev queries run with \vanilla{} and \systemlong{} on a TITAN V GPU. \Vanilla{} is overwhelmingly bottlenecked with the cost of index lookup and decompression, a challenge that \system{} addresses.}
    
    \label{fig:msmarco_latency_breakdown}
\end{figure}
\begin{figure*}
    \centering
     \includegraphics[scale=0.4]{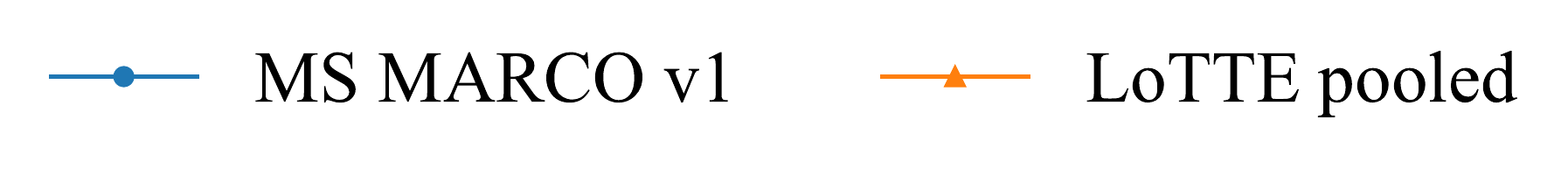} \\
    \begin{subfigure}[b]{0.33\linewidth}
    \centering
    \includegraphics[scale=0.37]{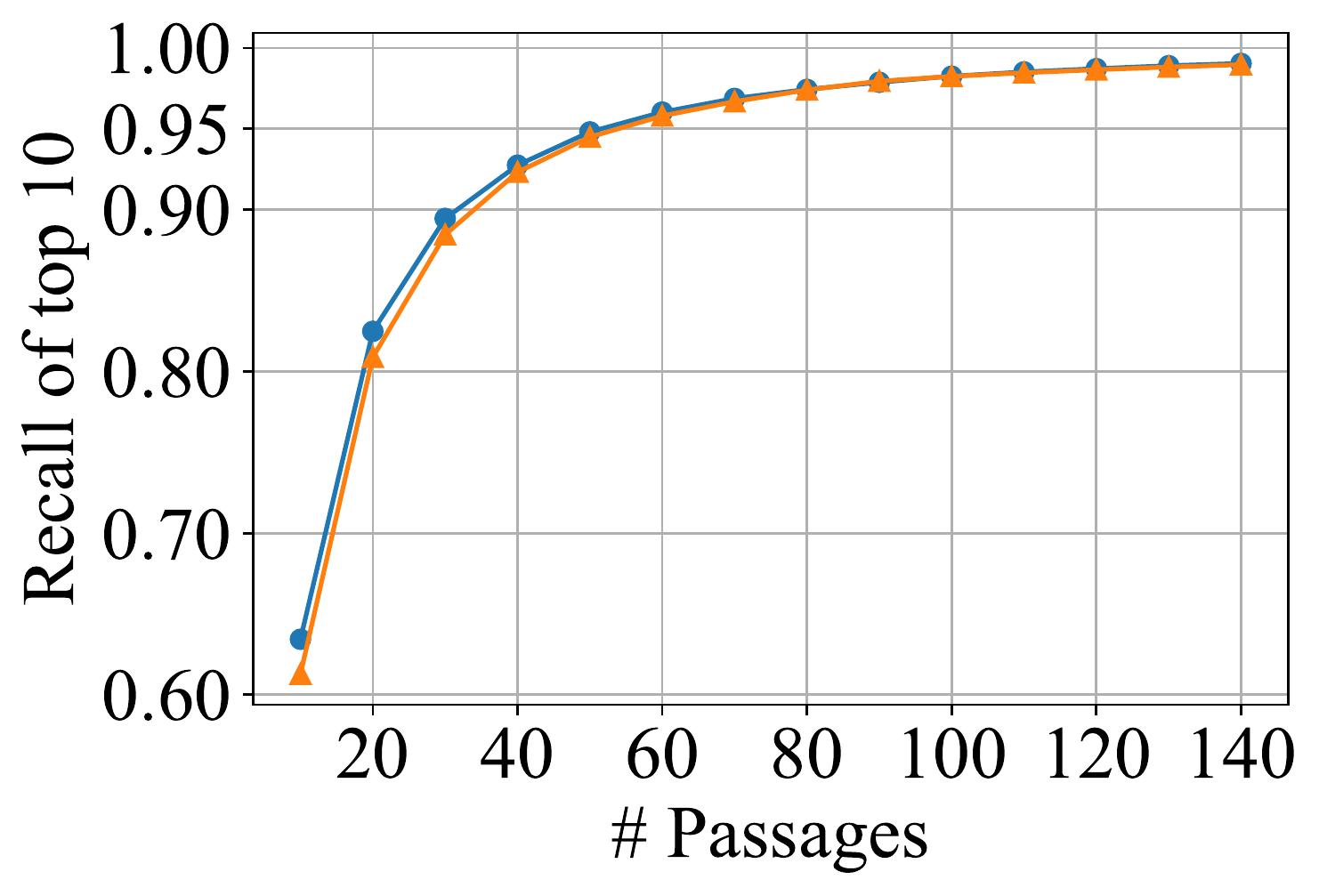}
    \caption{$k=10$}
    \label{fig:self_recall_k=10}
    \end{subfigure}
    \begin{subfigure}[b]{0.33\linewidth}
    \centering
    \includegraphics[scale=0.37]{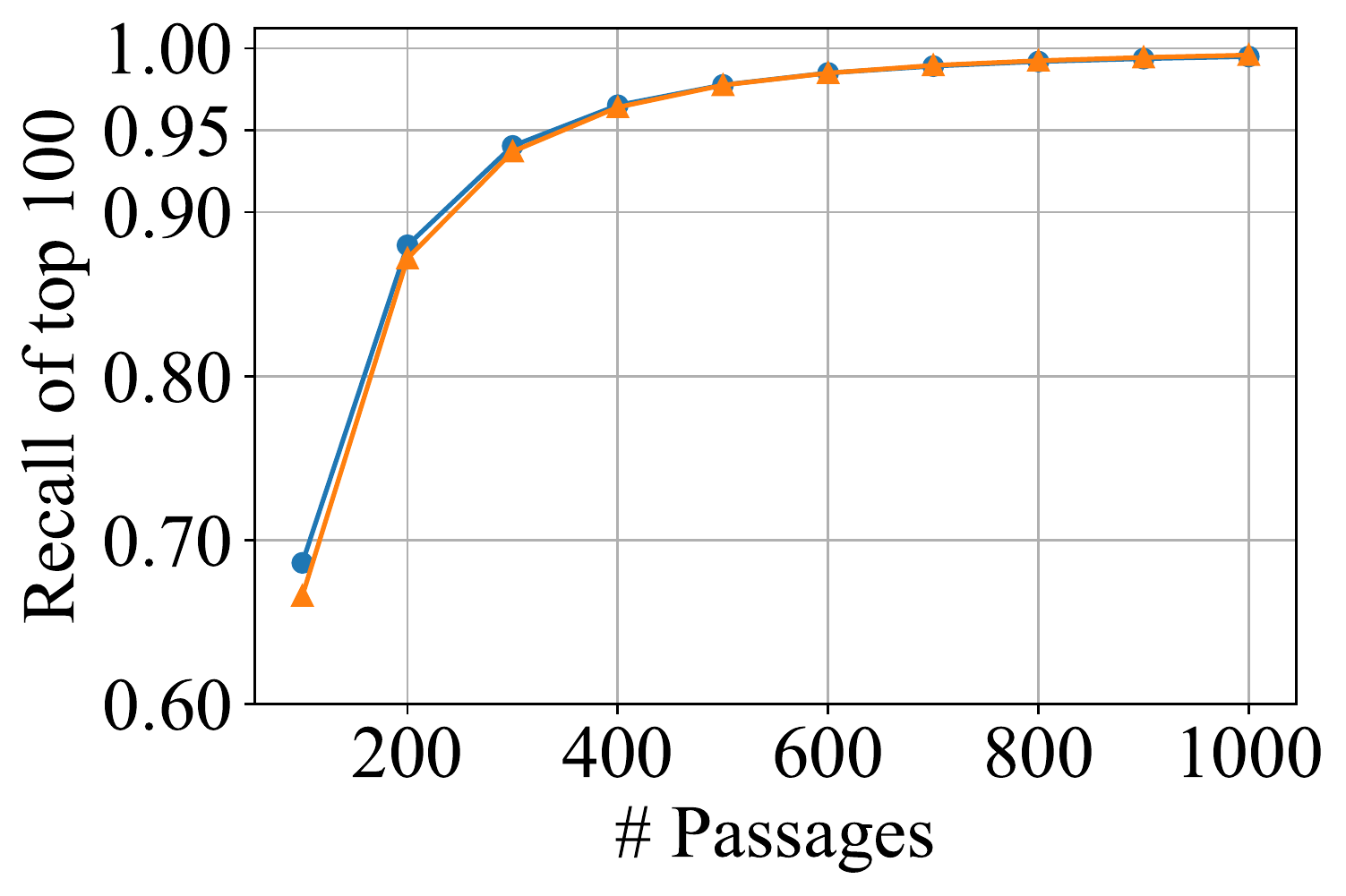}
    \caption{$k=100$}
    \label{fig:self_recall_k=100}
    \end{subfigure}
    \begin{subfigure}[b]{0.33\linewidth}
    \centering
    \includegraphics[scale=0.37]{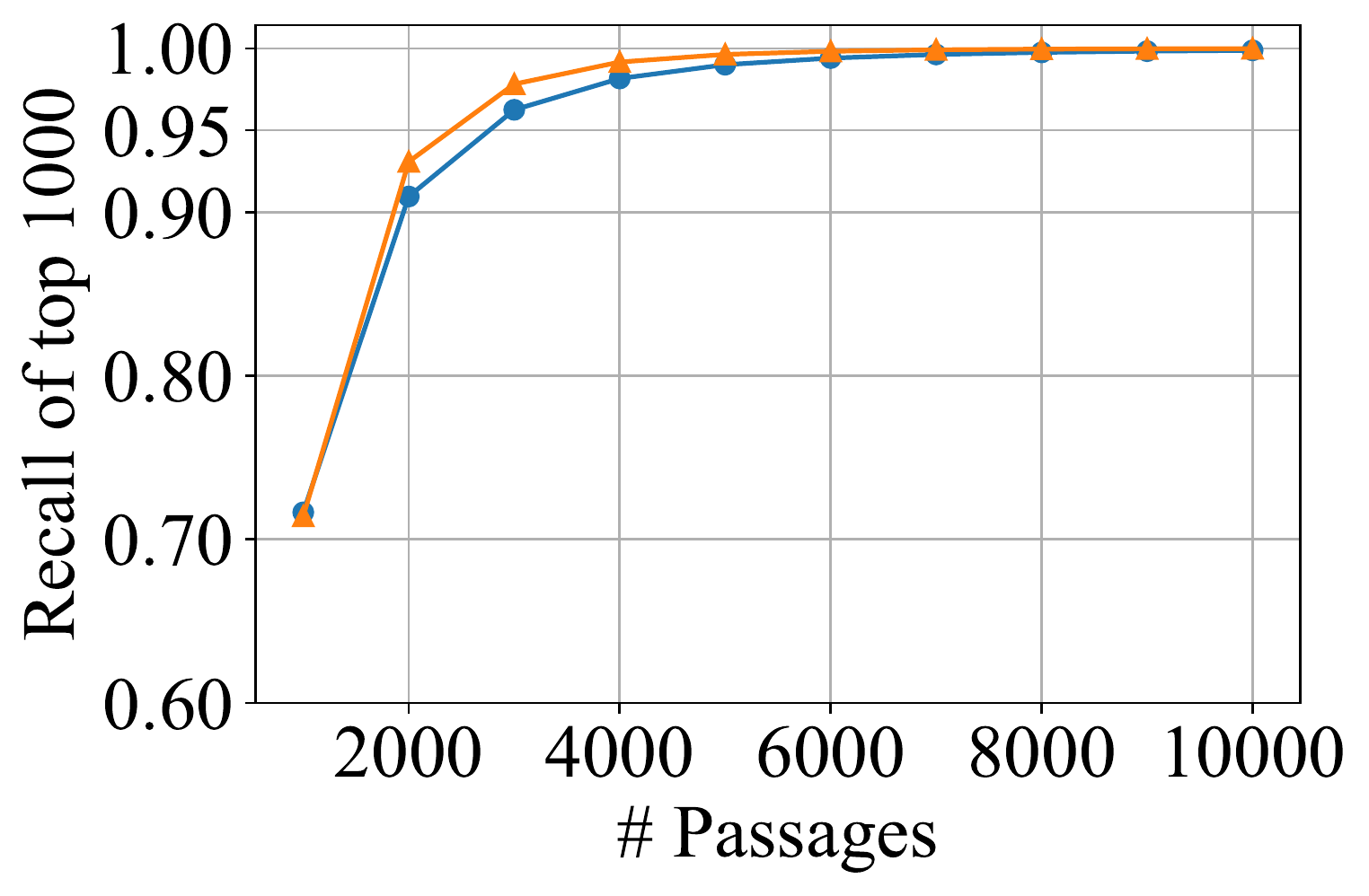}
    \caption{$k=1000$}
    \label{fig:self_recall_k=1000}
    \end{subfigure}
    \caption{Recall of passages retrieved by a centroid-only version of ColBERTv2 with respect to the top $k$ passages retrieved by \vanilla{}. Centroids alone can identify virtually all of the top-$k$ passages retrieved with the full ColBERTv2 pipeline, within $10 \cdot k$ or fewer candidates, motivating our centroid interaction strategy.}
    \label{fig:self_recall}
\end{figure*}

\subsection{Pruning for Sparse and Dense Retrieval} \label{subsection:background:pruning}

For sparse retrieval models, traditional IR has a wealth of work on fast strategies for skipping documents for top-$k$ search. Strategies often keep metadata like term score upper bounds to skip lower-scoring candidates and most follow a Document-At-A-Time (DAAT) scoring approach~\cite{turtle1995query,broder2003efficient,ding2011faster,dimopoulos2013optimizing,mallia2017faster, khattab2020finding}. Refer to \citet{tonellotto2018efficient} for a detailed treatment of recent methods. A key difference to our settings is that these all strategies expect a set of precomputed scores (particularly, useful upper bounds on every term--document pair), whereas with late interaction the term--document interaction (i.e., the MaxSim score) is only known at query time after a matrix-vector multiplication. Our observations about the utility of centroids for accelerating late interaction successfully moves the problem closer to classical IR, but poses the challenge that the query-to-centroid scores are only known at query time.

For dense retrieval models that use single-vector representations, approximate $k$-nearest neighbor (ANN) search is a well-studied problem~\cite{jegou2010product,johnson2019billion,abuzaid2019index,malkov2018efficient}. Our focus extends such work from a single vector to the late interaction of two matrices. %

\section{Analysis of ColBERTv2 Retrieval} \label{section:analysis}

We begin by a preliminary investigation of the latency~(\secref{subsection:analysis:latency_breakdown}) and scoring patterns~(\secref{subsection:analysis:centroids}) of ColBERTv2 retrieval that motivates our work on \system{}. To make this section self-contained, \secref{subsection:analysis:review} reviews the modeling, storage, and supervision of ColBERTv2.

\subsection{Modeling, Storage, and Retrieval} \label{subsection:analysis:review}

\system{} optimizes retrieval for models using the late interaction architecture of ColBERT, which includes systems like ColBERTv2, Baleen~\cite{khattab2021baleen}, Hindsight~\cite{paranjape2021hindsight}, and DrDecr~\cite{li2021learning}, among others. As depicted in Figure~\ref{fig:ColBERT}, a Transformer encodes queries and passages independently into vectors at the token level. For scalability, passage representations are pre-computed offline. At search time, the similarity between a query $q$ and a passage $d$ is computed as the summation of ``MaxSim'' operations, namely, the largest cosine similarity between each vector in the query matrix and all of the passage vectors:
\begin{equation}
\label{eq:scorer}
S_{q,d} = \sum_{i=1}^{|Q|} \max_{j=1}^{|D|} Q_{i} \cdot D_{j}^{T}
\end{equation}
where $Q$ and $D$ are the matrix representations of the query and passage, respectively. In doing so, this scoring function aligns each query token with the ``most similar'' passage token and estimates relevance as the sum of these term-level scores. Refer to \citet{khattab2020colbert} for a more complete discussion of late interaction.

For storing the passage representations, we adopt the ColBERTv2 residual compression strategy, which reduces the index size by up to an order of magnitude over naive storage of late-interaction embeddings as vectors of 16-bit floating-point numbers. Instead, ColBERTv2's compression strategy efficiently clusters all token-level vectors and encodes each vector using the ID of its nearest cluster centroid as well as a quantized residual vector, wherein each dimension is 1- or 2-bit encoding of the delta between the centroid and the original uncompressed vector. Decompressing a vector requires locating its centroid ID, encoded using 4 bytes, and its residual, which consume 16 or 32 bytes for 1- or 2-bit residuals, assuming the default 128-dimensional vectors.

While we adopt ColBERTv2's compression, we improve its retrieval strategy. We refer to the original retrieval strategy as ``vanilla'' ColBERTv2 retrieval. We refer to \citet{santhanam2021colbertv2} for details of compression and retrieval in ColBERTv2.

\subsection{ColBERTv2 Latency Breakdown} \label{subsection:analysis:latency_breakdown}

Figure~\ref{fig:msmarco_latency_breakdown} presents a breakdown of query latency on MS MARCO Passage Ranking (v1) on a GPU, showing results for \vanilla{} (Figure~\ref{fig:msmarco_baseline_latency_breakdown}) against the new \systemlong{} (Figure~\ref{fig:msmarco_plaid_latency_breakdown}). Latency is divided between query encoding, candidate generation, index lookups (i.e., to gather the compressed vector representations for candidate passages), residual decompression, and finally scoring (i.e., the final MaxSim computations).

For \vanilla{}, index lookup and residual decompression are overwhelming bottlenecks. Gathering vectors from the index is expensive because it consumes significant memory bandwidth: each vector in this setting is encoded with a 4-bit centroid ID and 32-byte residuals, each passage contains tens of vectors, and there can be up to $2^{16}$ candidate passages. Moreover, index lookup in \vanilla{} also constructs padded tensors on the fly to deal with the variable length of passages. Decompression of residuals is comprised of several non-trivial operations such as unpacking bits and computing large sums, which can be expensive when ColBERTv2 produces a large initial candidate set (\textasciitilde10-40k passages) as is the case for MS MARCO v1. While it is possible to use a smaller candidate set, doing so reduces recall (\secref{section:evaluation}).

\begin{figure}
    \centering
    \includegraphics[scale=0.4]{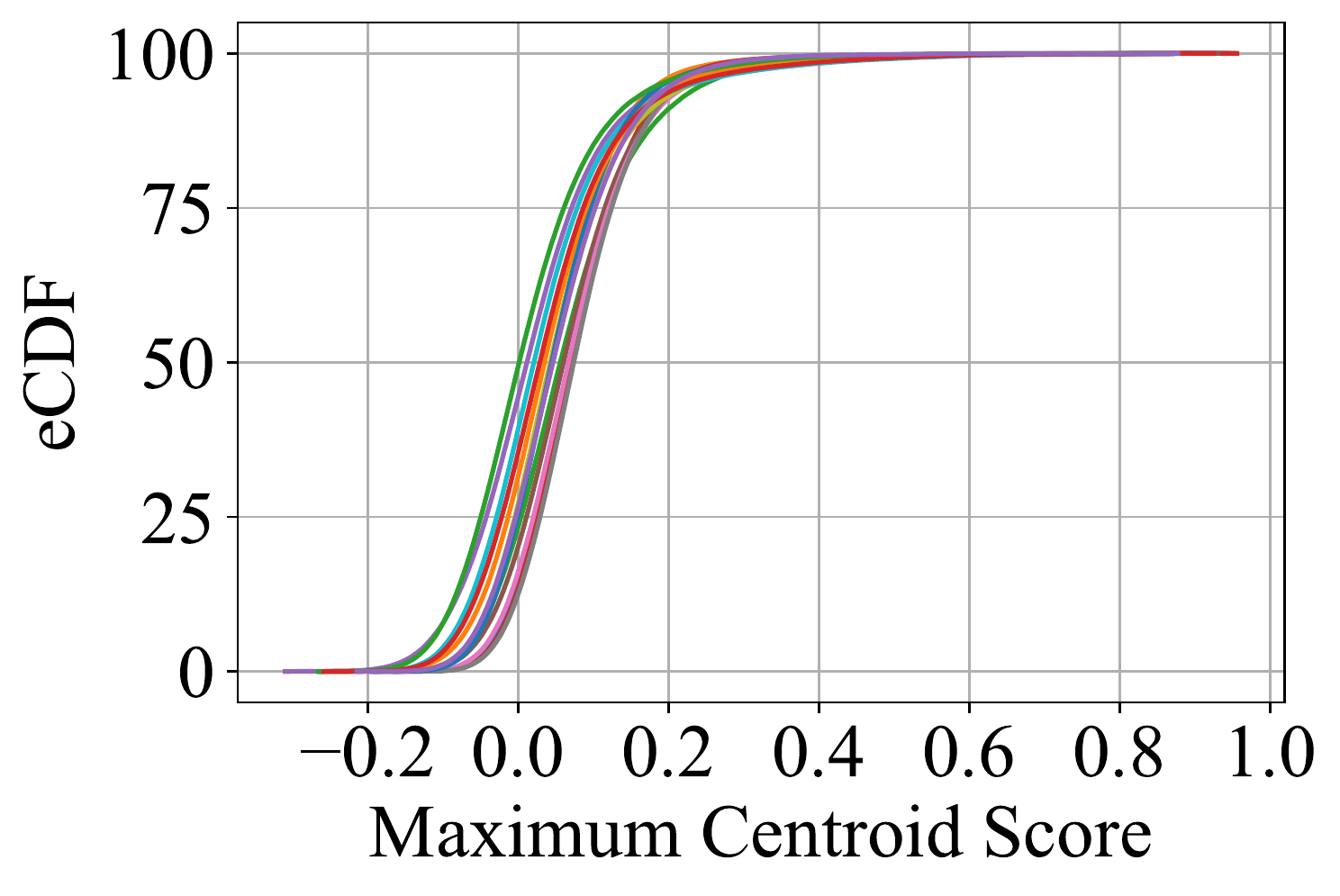}
    \caption{Centroid score distribution for each query among a random sample of 15 MS MARCO v1 dev queries evaluated with ColBERTv2.}
    \label{fig:centroid_score_distribution}
\end{figure}

\subsection{Centroids Alone Identify Strong Candidates} \label{subsection:analysis:centroids}

This breakdown in Figure~\ref{fig:msmarco_plaid_latency_breakdown} demonstrates that exhaustively scoring a large number of candidates passages, particularly gathering and decompressing their residuals, can amount to a considerable cost. Whereas ColBERTv2~\cite{santhanam2021colbertv2} exploits centroids to reduce the space footprint, our work demonstrates that the centroids \textit{can also accelerate search}, while maintaining quality, by serving as proxies for the passage embeddings. Because of this, we can skip low-scoring passages without having to look up or decompress their residuals, adding some additional candidate generation overhead to achieve substantial savings in the subsequent stages (Figure~\ref{fig:msmarco_plaid_latency_breakdown}).

Effectively, we hypothesize that \textit{centroid-only retrieval} can find the high-scoring passages otherwise retrieved by \vanilla{}. We test this hypothesis by comparing the top-$k$ passages retrieved by \vanilla{} to a modified implementation that conducts retrieval using only the centroids and no residuals. We present the results in Figure~\ref{fig:self_recall}. At $k \in \{10, 100, 1000\}$, the figure plots the average recall of the top-$k$ passages of \vanilla{} within the passages retrieved by centroid-only ColBERTv2 at various depths. In other words, we report the fraction of the top-$k$ passages of \vanilla{} that appear within the top-$k'$ passages of centroid-only ColBERTv2, for $k' \geq k$.

The results support our hypothesis, both in domain for MS MARCO v1 and out of domain using the LoTTE Pooled (dev) search queries~\cite{santhanam2021colbertv2}. For instance, if we retrieve $10 \cdot k$ passages using only centroids, those $10 \cdot k$ passages still contain 99+\% of the top $k$ passages retrieved by the \vanilla{} full pipeline.

\subsection{Not All Centroids Are Important Per Query} \label{subsection:analysis:centroids2}

We further hypothesize that for a given query a small subset of the passage embedding clusters tend to be far more important than others in determining relevance. If this were in fact the case, then we could prioritize computation over these highly weighted centroids and discard the rest since we know they will not contribute significantly to the final ranking. We test this theory by randomly sampling 15 MS MARCO v1 queries and plotting an empirical CDF of each centroid's maximum relevance score observed across all query tokens, as shown in Figure~\ref{fig:centroid_score_distribution}. We do find that there is a small tail of highly weighted centroids whose relevance scores have far higher magnitude than all other centroids. While not shown in Figure~\ref{fig:centroid_score_distribution}, we also repeated this experiment with LoTTE pooled queries and found a very similar score distribution.

\section{\system{}} \label{section:centroid_scoring_pipeline}

\begin{figure*}[ht]
    \centering
    \includegraphics[scale=0.35]{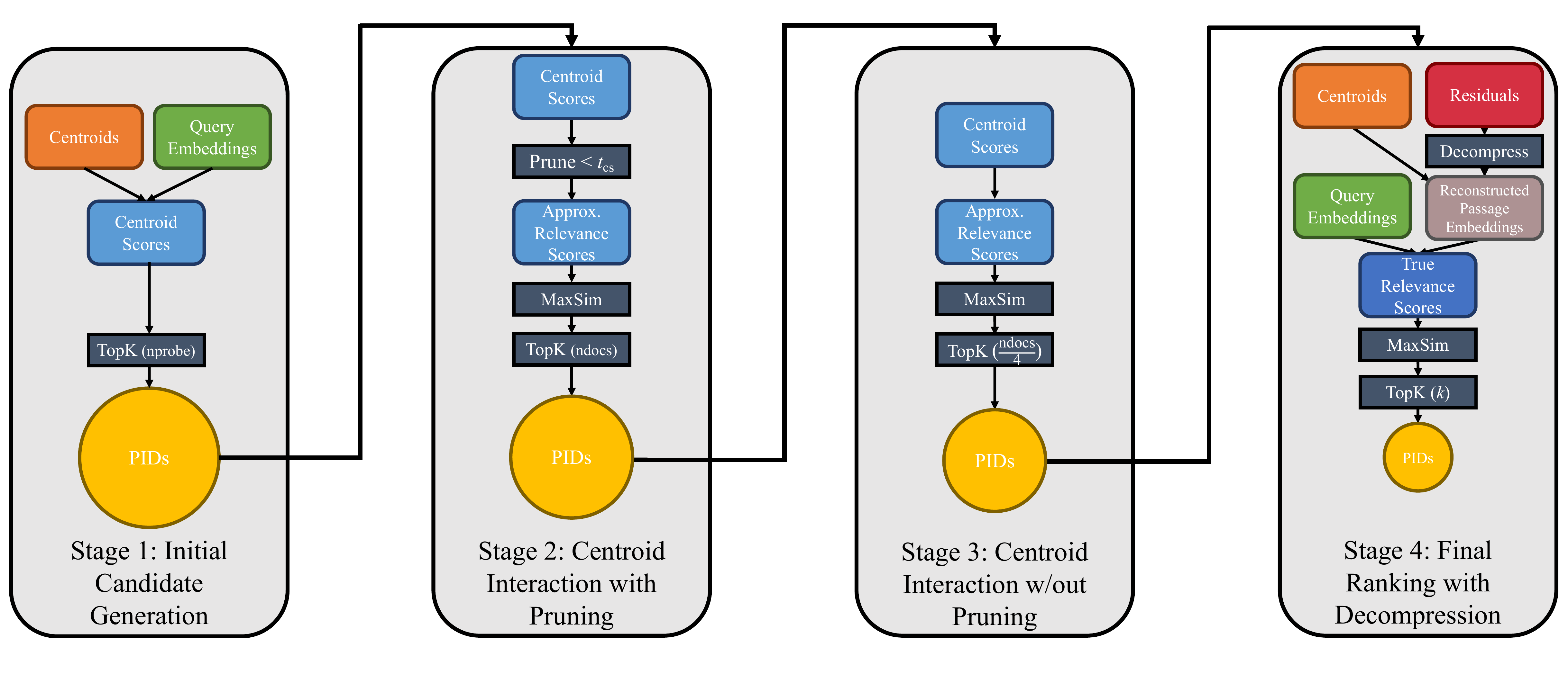}
    \caption{The \system{} scoring pipeline. The first stage generates an initial set of candidate passages using the centroids. Next the second and third stages leverage centroid pruning and centroid interaction respectively to refine the candidate set. Then the last stage performs full residual decompression to obtain the final passage ranking. We use the hyperparameter \texttt{ndocs} to specify the number of candidates returned by Stage 2, and in our experiments we have Stage 3 output $\frac{\texttt{ndocs}}{4}$ passages.}
    \label{fig:scoring_pipeline}
\end{figure*}

Figure~\ref{fig:scoring_pipeline} illustrates the \system{} scoring pipeline, which consists of multiple consecutive stages for retrieval, filtering, and ranking. The first stage produces an initial candidate set by computing relevance scores for each centroid with respect to the query embeddings. In the intermediate stages, \system{} uses the novel techniques of \emph{centroid interaction} and \emph{centroid pruning} to aggressively yet effectively filter the candidate passages. Finally, \system{} ranks the final candidate set using fully reconstructed passage embeddings. We discuss each of these modules in more depth as follows.

\subsection{Candidate Generation}

Given the query embedding matrix $Q$ and the list of centroid vectors $C$ in the index, \system{} computes the token-level query--centroid relevance scores $S_{c,q}$ as a matrix multiplication:
\begin{equation} \label{eq:centroid_scores}
    S_{c,q} = C \cdot Q^{T} \\
\end{equation} 
and then identifies the passages ``close'' to the top-$t$ centroids per query token as the initial candidate set. A passage is close to a centroid \textit{iff} one or more of its tokens are assigned to that centroid by $k$-means clustering during indexing. This value $t$ is referred to as \texttt{nprobe} in \vanilla{} and we retain that terminology in \systemlong{}.  

The initial candidate generation in \systemlong{} differs from the corresponding \vanilla{} stage in two key aspects. First, while \vanilla{} saves an inverted list mapping centroids to their corresponding embedding IDs, \systemlong{} instead structures the inverted list as a map from centroids to the corresponding \textit{unique passage IDs}. Storing passage IDs is advantageous over storing embedding IDs since there are far fewer passages than embeddings, meaning the inverted list has to store less information overall. This also enables \systemlong{} to use 32-bit integers in the inverted list rather than potentially 64-bit longs.\footnote{This assumes no more than $\le 2^{32}$ (4 billion) passages in the corpus, but this limit is 30$\times$ larger than even MS MARCO v2~\cite{craswell2022overview}.} In practice, this translates to a space savings of 2.7$\times$ in the MS MARCO v2~\cite{craswell2022overview} inverted list (71 GB to 27 GB, with 140M passages).

Second, and relatedly, if the initial candidate set was too large (as specified by the \texttt{ncandidates} hyperparameter) \vanilla{} would prune it by scoring and ranking a subset of the candidate embedding vectors---in particular, the embeddings listed within the vanilla mapping from centroid IDs to embedding IDs---with full residual decompression, which is quite costly as we discuss in \secref{subsection:analysis:latency_breakdown}. In contrast, \systemlong{} does not impose any limit on the initial candidate size because the subsequent stages can cheaply filter the candidate passages with centroid interaction and pruning.

\subsection{Centroid Interaction}

Centroid interaction cheaply approximates per-passage relevance by substituting each token's embedding vector with its nearest centroid in the standard MaxSim formulation. By applying centroid interaction as an additional filtering stage, the scoring pipeline can skip the expensive embedding reconstruction process for a large fraction of the candidate passages. This results in significantly faster end-to-end retrieval. Intuitively, centroid interaction enables \system{} to emulate traditional bag-of-words retrieval wherein the centroid relevance scores take the role of the term relevance scores used in systems like BM25. However, because of its vector representations (of the query in particular), \system{} computes the centroid relevance scores at query time in contrast to the more traditional pre-computed term relevance scores. 

The procedure works as follows. Recall that $S_{c,q}$ from Equation~\ref{eq:centroid_scores} stores the relevance scores for each centroid with respect to the query tokens. Suppose $I$ is the list of the centroid indices mapped to each of the tokens in the candidate set. Furthermore, let $S_{c,q}[i]$ denote the $i$-th row of $S_{c,q}$. Then \system{} constructs the centroid-based approximate scores $\tilde{D}$ as 
\begin{equation}
    \tilde{D} = \begin{bmatrix}
         S_{c,q}[I_{1}] \\
         S_{c,q}[I_{2}] \\
         \cdots \\
         S_{c,q}[I_{|\tilde{D}|}]
    \end{bmatrix}
\end{equation}
Then to rank the candidate passages using $\tilde{D}$, \system{} computes the MaxSim scores $S_{\tilde{D}}$ as
\begin{equation}
    S_{\tilde{D}} = \sum_{i}^{|Q|} \max_{j=1}^{|\tilde{D}|} \tilde{D}_{i,j}  
\end{equation}
The top $k$ most relevant passages drawn from $S_{\tilde{D}}$ serve as the filtered candidate passage set.

\system{} includes optimized kernels to efficiently deploy centroid interaction (and more generally MaxSim operations); we discuss these in \secref{subsection:plaid:padding_free_maxsim}.

\subsection{Centroid Pruning}

As an additional optimization, \system{} leverages the observation from \S\ref{subsection:analysis:centroids} to first prune low-magnitude centroid scores before constructing $\tilde{D}$. In this filtering phase \system{} will only score tokens whose maximum corresponding centroid score meets the given threshold $t_{cs}$. Concretely, $\tilde{D}$ will only be comprised of tokens whose corresponding centroid (suppose centroid $i$) meets the following condition:
\begin{equation}
    \max_{j=1}^{|Q|} S_{c,q_{i, j}} \ge t_{cs}
\end{equation}

We introduce the hyperparameter \texttt{ndocs} to refer to the number of candidate documents selected by Stage 2. We then found empirically that choosing $\frac{\texttt{ndocs}}{4}$ candidates from Stage 3 produced good results; we use this heuristic for all the results presented in \S\ref{section:evaluation}.

\subsection{Scoring}

As in \vanilla{}, \system{} will reconstruct the original embeddings of the final candidate passage set via residual decompression and rank these using MaxSim. Let $D$ be the reconstructed embedding vectors for the final candidate set after decompression. Then the final scores $S_{q,d}$ are computed using Equation~\ref{eq:scorer}.

Section \secref{subsection:plaid:padding_free_maxsim} discusses fast kernels for accelerating the MaxSim and decompression operations.

\subsection{Fast Kernels: Padding-Free MaxSim \& Optimized Decompression}
\label{section:implementation} \label{subsection:plaid:padding_free_maxsim}

Figure~\ref{fig:msmarco_baseline_latency_breakdown} shows that index lookup operations are a large source of overhead for \vanilla. One reason these lookups are expensive is that they require reshaping and padding the 2D index tensors with an extra dimension representing the maximum passage length. The resulting 3D tensors facilitate batched MaxSim operations over ragged lists of token vectors.  To avoid this padding, we instead implement custom C++ code that directly computes the MaxSim scores over the \textit{packed} 2D index tensors (i.e., one where many 2D sub-tensors of various lengths are concatenated along the same dimension). Our kernel loops over each passage's corresponding token vectors to compute the per-passage maximum scores with respect to each query token and then sums the per-passage maximum scores across all query tokens. This design is trivial to parallelize across passages, and also enables $O(|Q|)$ per-thread memory usage by allocating a single output vector to store the maximum scores per query token and repeatedly updating this vector in-place. In contrast, the padding-based approach requires $O(|D| \cdot |Q|)$ space. We have incorporated this design into optimized implementations of centroid interaction as well as the final MaxSim operation (stage 4 in Figure~\ref{fig:scoring_pipeline}). \system{} only implements these kernels for CPU execution. Adding corresponding GPU kernels remains future work.

\label{subsection:plaid:decompression}

ColBERTv2's residual decompression scheme computes a list of centroid vectors, determines a fixed set of $2^{b}$ possible deltas from these centroids, and then stores the index into the set of deltas corresponding to each embedding vector. In particular, each compressed 8-bit value stores $\frac{8}{b}$ indices in the range $[0, 2^{b})$. ColBERTv2 incurs significant overhead due to residual decompression, as shown in Figure~\ref{fig:msmarco_baseline_latency_breakdown}. This is partially due to the na\"ive decompression implementation, which required explicitly unpacking bits from the compressed representation and performing expensive bit shift and sum operations to recover the original values. Instead, \system{} pre-computes all $2^{8}$ possible lists of indices encoded by an 8-bit packed value. These outputs are stored in a lookup table so that the decompression function can simply retrieve the indices from the table rather than manually unpacking the bits. We include optimized implementations of this lookup-based decompression for both CPU and GPU execution. The GPU implementation uses a custom CUDA kernel that allocates a separate thread to decompress each individual byte in the compressed residual tensor (the thread block size is computed as $\frac{b \cdot d}{8}$ for $d$-dimensional embedding vectors). The CPU implementation instead parallelizes decompression at the granularity of individual passages.

\section{Evaluation} \label{section:evaluation}

Our evaluation seeks to answer the following research questions:
\begin{enumerate}
    \item How does \system{} affect end-to-end latency and retrieval quality across IR benchmarks? (\S\ref{subsection:eval:end_to_end})
    \item How much do each of \system{}'s optimizations contribute to the performance speedups? (\S\ref{subsection:eval:ablation})
    \item How well does \system{} scale with respect to the corpus size and the parallelism degree? (\S\ref{subsection:eval:scalability})
\end{enumerate}

\subsection{Setup} \label{subsection:eval:setup}

\begin{table}[]
\centering
\resizebox{\columnwidth}{!}{%
\begin{tabular}{@{}lrrrrr@{}}
\toprule
\multicolumn{1}{c}{\multirow{2}{*}{Dataset}} & \multicolumn{1}{c}{\multirow{2}{*}{\# Passages}} & \multicolumn{1}{c}{\multirow{2}{*}{\# Tokens}} & \multicolumn{1}{c}{\multirow{2}{*}{\# Queries}}  & \multicolumn{2}{c}{\begin{tabular}[c]{@{}c@{}}ColBERTv2\\Index Size (GiB)\end{tabular}} \\ \cmidrule{5-6}
\multicolumn{1}{c}{} & \multicolumn{1}{c}{} & \multicolumn{1}{c}{} & \multicolumn{1}{c}{}  & \multicolumn{1}{c}{Vanilla} & \multicolumn{1}{c}{\system{}} \\ \midrule
MS MARCO v1~\cite{nguyen2016ms}                     & 8.8M                            & 597.9M                    & 6980        & 24.6      & 21.6                 \\
Wikipedia~\cite{karpukhin2020dense}                 & 21.0M                           & 2.6B                      & 8757        & 105.2          &   92.0          \\
LoTTE pooled~\cite{santhanam2021colbertv2}          & 2.4M                            & 339.4M                    & 2931        & 14.0      & 12.3              \\
MS MARCO v2~\cite{craswell2022overview}             & 138.4M                          & 9.4B                      & 3903        & 246.0     & 202.2            \\ \bottomrule
\end{tabular}
}
\caption{List of benchmarks used for evaluation with relevant statistics.}
\label{table:eval_datasets}
\end{table}
\begin{table}[]
\begin{tabular}{@{}crrr@{}}
\toprule
$k$    & \multicolumn{1}{c}{\texttt{nprobe}} & \multicolumn{1}{c}{$t_{cs}$} & \multicolumn{1}{c}{\texttt{ndocs}} \\ \midrule
10   & 1                          & 0.5                                            & 256                       \\
100  & 2                          & 0.45                                           & 1024                      \\
1000 & 4                          & 0.4                                            & 4096                      \\ \bottomrule
\end{tabular}
\caption{\system{} hyperparameter configuration.}
\label{table:plaid_parameters}
\end{table}

\paragraph{\system{} Implementation} The \system{} engine subsumes centroid interaction as well as optimizations for residual decompression. We implement \system{} modularly as an extension to ColBERTv2's PyTorch-based implementation, particularly its search components. For CPU execution, we implement the centroid interaction and decompression operations entirely in multithreaded C++ code. For GPUs, we implement centroid interaction in PyTorch and provide a CUDA kernel for fast decompression. Overall, \system{} constitutes roughly 300 lines of additional Python code and 700 lines of C++.

\paragraph{Datasets} Our evaluation includes results from four different IR benchmarks, as listed in Table~\ref{table:eval_datasets}. We perform in-domain evaluation on the MS MARCO v1 and Wikipedia Open QA benchmarks, with retrievers trained specifically for these tasks, and out-of-domain evaluation on the StackExchange-based LoTTE~\citet{santhanam2021colbertv2} and the TREC 2021 Deep Learning Track~\cite{craswell2022overview} MS MARCO v2 benchmarks, with the ColBERTv2 retriever~\cite{santhanam2021colbertv2} trained on MS MARCO v1. For evaluation on Wikipedia we use the December 2018 dump~\cite{karpukhin2020dense} with queries from the NaturalQuestions (NQ) dataset~\cite{kwiatkowski2019natural}. Our LoTTE~\cite{santhanam2021colbertv2} evaluation uses the ``pooled'' dev dataset with ``search''-style queries. For MS MARCO v2, we use the augmented passage version of the data~\cite{lin2022passage} and include passage titles while ignoring headings. As we evaluate several configurations of the models, all of our evaluation is performed using development set queries.

\paragraph{Systems and hyperparameters} We report results for several systems for end-to-end results: \vanilla{} and \systemlong{} as well as ColBERT (v1)~\cite{khattab2020colbert}, BM25~\cite{robertson1995okapi}, SPLADEv2~\cite{formal2021spladev2}, and DPR~\cite{karpukhin2020dense}. For \vanilla, we use the specific hyperparameters reported in the ColBERTv2 paper for each benchmark dataset. We indicate these in the result tables with \texttt{p} (\texttt{nprobe}) and \texttt{c} (\texttt{ncandidates}). For \systemlong{}, we evaluate
three different settings: $k=10$, $k=100$, and $k=1000$. The $k$ parameter controls the final number of scored documents as well as the retrieval hyperparameters described in \secref{section:centroid_scoring_pipeline}. Table~\ref{table:plaid_parameters} lists these hyperparameter configurations for each $k$ setting. We find empirically that ranking $\frac{\texttt{ndocs}}{4}$ documents for the final scoring stage produces strong results. For both \vanilla{} and \systemlong{}, we compress all datasets to 2 bits per dimension, with the exception of MS MARCO v2 where we compress to 1 bit.

\paragraph{Hardware} We conduct all experiments on servers with 28 Intel Xeon Gold 6132 2.6 GHz CPU cores (2 threads per core for a total of 56 threads) and 4 NVIDIA TITAN V GPUs each. Every server has two NUMA sockets with roughly 92 ns intra-socket memory latency, 142 ns inter-socket memory latency, 72 GBps intra-socket memory bandwidth, and 33 GBps inter-socket memory bandwidth. Each TITAN V GPU has 12 GB of high-bandwidth memory. 

\paragraph{Latency measurements} When measuring latency for end-to-end results, we compute the average latency of all queries (see Table~\ref{table:eval_datasets} for query totals), and then report the minimum average latency across 3 trials. For other results we describe the specific measurement procedure in the relevant section. We discard the query encoding latency for neural models (ColBERTv1~\cite{khattab2020colbert}, \vanilla{}~\cite{santhanam2021colbertv2}, \systemlong{}, and SPLADEv2~\cite{formal2021spladev2}) following \citet{mackenzie2021wacky}; prior work has shown that the cost of running the BERT model can be made negligible with standard techniques such as quantization, distillation, etc.~\cite{bergum2021pretrained}. We measure latency on an otherwise idle machine. We prepend commands with \texttt{numactl -{}-membind 0} to ensure intra-socket I/O operations. We do not do this for MS MARCO v2, since its large index may require both NUMA nodes. For GPU results we allow full usage of all 56 threads, but for CPU-only results we restrict usage to either 1 or 8 threads using \texttt{torch.set\_num\_threads}. For non-ColBERT systems we use the single-threaded latency numbers reported by \citet{mackenzie2021wacky}. Note that these numbers were measured on a different hardware setup and using a different implementation and are therefore simply meant to establish \systemlong{}'s competitive performance rather than serving as absolute comparisons.

\subsection{End-to-end Results} \label{subsection:eval:end_to_end}

\begin{table}[ht]
\centering
\resizebox{\columnwidth}{!}{%
\setlength{\tabcolsep}{1.5pt}
\begin{tabular}{@{}lrrrrrr@{}}
\toprule
\multicolumn{1}{c}{\multirow{2}{*}{System}} & \multicolumn{1}{c}{\multirow{2}{*}{MRR@10}}  & \multicolumn{1}{c}{\multirow{2}{*}{R@100}} & \multicolumn{1}{c}{\multirow{2}{*}{R@1k}}  &
\multicolumn{3}{c}{Latency (ms)}                                                \\ \cmidrule(l){5-7} 
\multicolumn{1}{c}{}                            & \multicolumn{1}{c}{}      &   \multicolumn{1}{c}{}    & \multicolumn{1}{c}{}  & \multicolumn{1}{c}{1-CPU} & \multicolumn{1}{c}{8-CPU}   & \multicolumn{1}{c}{GPU} \\ \midrule 
BM25 (PISA~\cite{mallia2019pisa}; $k=1000$)                                 & 18.7*                     & -                         & -                     & 8.3*       & -               & -                       \\
SPLADEv2 (PISA; $k=1000$)                             & 36.8*                     & -                         & 97.9*                 & 220.3*     & -               & -                       \\
\midrule
ColBERTv1                                    & 36.1                      & 87.3                      & 95.2                  & -         & -               & 54.3                    \\
\Vanilla (\texttt{p}=2, \texttt{c}=$2^{13}$)   	& 39.7                      & 90.4                      & 96.6                  & 3485.1    & 921.8           & 53.4                    \\
\Vanilla (\texttt{p}=4, \texttt{c}=$2^{16}$)   	& 39.7                      & 91.4                      & 98.3                  & -         & 4568.5          & 259.6                   \\
\midrule
\systemlong{} ($k=10$)                        	& 39.4                      & -                         & -                     & 185.5     & 31.5            & 11.5                   \\
\systemlong{} ($k=100$)                       	& 39.8                      & 90.6                      & -                     & 222.3     & 52.9            & 20.2                   \\
\systemlong{} ($k=1000$)                      	& 39.8                      & 91.3                      & 97.5                  & 352.3     & 101.3           & 38.4                   \\  \bottomrule
\end{tabular}
}
\caption{End-to-end in-domain evaluation on the MS MARCO v1 benchmark. Numbers marked with an asterisk are copied from \citet{formal2021splade} for SPLADEv2 quality and \citet{mackenzie2021wacky} for latencies.}
\label{table:msmarco_end_to_end_results}
\end{table}

\begin{table}[h]
\centering
\resizebox{\columnwidth}{!}{%
\setlength{\tabcolsep}{2pt}
\begin{tabular}{@{}lrrrr@{}}
\toprule
\multicolumn{1}{c}{\multirow{2}{*}{System}} & \multicolumn{1}{c}{\multirow{2}{*}{Success@5}}  & \multicolumn{1}{c}{\multirow{2}{*}{Success@100}} &
\multicolumn{2}{c}{Latency (ms)}                                                \\ \cmidrule(l){4-5} 
\multicolumn{1}{c}{}                            & \multicolumn{1}{c}{}  & \multicolumn{1}{c}{}      & \multicolumn{1}{c}{CPU (8)}   & \multicolumn{1}{c}{GPU} \\ \midrule 
DPR  &  66.8                 & 85.0                      & -                   & -                  \\
ColBERT-QA Retrieval (uncompressed)  &  75.3                 & 89.2                      & -                   & -                  \\
\midrule
\multicolumn{5}{c}{ColBERT-QA~\cite{khattab2021relevance} Retriever with ColBERTv2~\cite{santhanam2021colbertv2} residual compression}               \\
\midrule
Vanilla ColBERT-QA Retrieval (\texttt{p}=4, \texttt{c}=$2^{15}$)    &  74.3                 & 89.0                      & 5077.9                    & 204.1                    \\
\system{} ColBERT-QA Retrieval ($k=10$)                        	&  73.3                 & -                         & 67.1                      & 13.6                    \\
\system{} ColBERT-QA Retrieval ($k=100$)                       	&  74.1                 & 88.0                      & 120.1                     & 26.9                   \\
\system{} ColBERT-QA Retrieval ($k=1000$)                      	&  74.4                 & 88.9                      & 228.4                     & 55.3                    \\  \bottomrule
\end{tabular}
}

\caption{End-to-end in-domain retrieval evaluation on the Wikipedia open-domain question answering benchmark. We use the NQ checkpoint of ColBERT-QA~\cite{khattab2021relevance}, and apply ColBERTv2 compression. We compare \vanilla{} retrieval against \systemlong{} retrieval. DPR results from \citet{karpukhin2020dense}. We refer to \citet{khattab2021relevance} for details on OpenQA retrieval evaluation.}
\label{table:wiki_end_to_end_results}
\end{table}

\begin{table}[h]
\centering
\footnotesize
\begin{tabular}{@{}lrrrr@{}}
\toprule
\multicolumn{1}{c}{\multirow{2}{*}{System}} & \multicolumn{1}{c}{\multirow{2}{*}{Success@5}}  & \multicolumn{1}{c}{\multirow{2}{*}{Success@100}} &
\multicolumn{2}{c}{Latency (ms)}                                                \\ \cmidrule(l){4-5} 
\multicolumn{1}{c}{}                            & \multicolumn{1}{c}{}  & \multicolumn{1}{c}{}          & \multicolumn{1}{c}{CPU (8)}   & \multicolumn{1}{c}{GPU} \\ \midrule 
BM25                                            & 47.8*                     & 77.6*                         & -                          & -                       \\
SPLADEv2                                        & 67.0*                     & 89.0*                         & -                          & -                       \\
\midrule
\Vanilla{} (\texttt{p}=2, \texttt{c}=$2^{13}$)  & 69.3                      & 90.3                          & 1508.4                    & 66.9                      \\
\systemlong{} ($k=10$)                          & 69.1                      & -                             & 35.5                      & 9.2                   \\
\systemlong{} ($k=100$)                         & 69.4                      & 89.9                          & 64.8                      & 17.4                     \\
\systemlong{} ($k=1000$)                        & 69.6                      & 90.5                          & 163.1                     & 27.3                   \\  \bottomrule
\end{tabular}
\caption{End-to-end out-of-domain evaluation on the (dev) pooled dataset of the LoTTE benchmark. Numbers marked with an asterisk were taken from \citet{santhanam2021colbertv2}.}
\label{table:lotte_end_to_end_results}
\end{table}

\begin{table}[ht]
\centering
\resizebox{\columnwidth}{!}{%
\setlength{\tabcolsep}{3pt}
\begin{tabular}{@{}lrrrrr@{}}
\toprule
\multicolumn{1}{c}{\multirow{2}{*}{System}} & \multicolumn{1}{c}{\multirow{2}{*}{MRR@100}}  & \multicolumn{1}{c}{\multirow{2}{*}{R@100}} & \multicolumn{1}{c}{\multirow{2}{*}{R@1k}}  &
\multicolumn{2}{c}{Latency (ms)}                                                \\ \cmidrule(l){5-6} 
\multicolumn{1}{c}{}                              & \multicolumn{1}{c}{}      &   \multicolumn{1}{c}{}    & \multicolumn{1}{c}{}  & \multicolumn{1}{c}{8-CPU}   & \multicolumn{1}{c}{GPU} \\ \midrule 
BM25 (Anserini~\cite{yang2018anserini}; Augmented)      &                 8.7      &        40.3               &         69.3          & -                    & -                   \\
\midrule
\Vanilla (\texttt{p}=4, \texttt{c}=$2^{16}$)      & 18.0                      & 68.2                      & 88.1                  & 5228.5                    & OOM                   \\
\systemlong{} ($k=10$)                            & -                         & -                         & -                     & 136.4                     & 47.1                   \\
\systemlong{} ($k=100$)                           & 17.9                      & 67.0                      & -                     & 181.9                     & 96.1                   \\
\systemlong{} ($k=1000$)                          & 18.0                      & 68.4                      & 85.7                  & 251.3                     & OOM                   \\  \bottomrule
\end{tabular}
}
\caption{End-to-end out-of-domain evaluation on the MS MARCO v2 benchmark. BM25 results from~\cite{anserini2022marcov2bm25}.}
\label{table:msmarcov2_results}
\end{table}

Table~\ref{table:msmarco_end_to_end_results} presents in-domain results for the MS MARCO v1 benchmark. We observe that in the most conservative setting ($k=1000$), \systemlong is able to match the MRR@10 and Recall@100 achieved by \vanilla while delivering speedups of 6.8$\times$ on GPU and 45$\times$ on CPU. For some minimal reduction in quality, \systemlong can further increase the speedups over \vanilla to 12.9--22.6$\times$ on GPU and 86.4--145$\times$ on CPU. \systemlong also achieves competitive latency compared to other systems (within 1.6$\times$ of SPLADEv2) while outperforming them on retrieval quality. 

We observe a similar trend with in-domain evaluation on the Wikipedia OpenQA benchmark as shown in Table~\ref{table:wiki_end_to_end_results}. \systemlong achieves speedups of 3.7$\times$ on GPU and 22$\times$ on CPU with no quality loss compared to \vanilla, and speedups of 7.6--15$\times$ on GPU and 42.3--75.7$\times$ on CPU with minimal quality loss. 

We confirm \system{} works well in out-of-domain settings, as well, as demonstrated by our results on the LoTTE ``pooled'' dataset. We see in Table~\ref{table:lotte_end_to_end_results} that \systemlong{} outperforms \vanilla{} by 2.5$\times$ on GPU and 9.2$\times$ on CPU with $k=1000$; furthermore, this setting actually improves quality compared to \vanilla{}. With some quality loss \systemlong{} can achieve speedups of 3.8--7.3$\times$ on GPU and 23.2--42.5$\times$ on CPU. Note that the CPU latencies achieved on LoTTE with \systemlong{} are larger than those achieved on MS MARCO v1 because the average LoTTE passage length is roughly 2$\times$ that of MS MARCO v1.

Finally, Table~\ref{table:msmarcov2_results} shows that \systemlong{} scales effectively to MS MARCO v2, which is a large-scale dataset with 138M passages and 9.4B tokens (approximately 16$\times$ bigger than MS MARCO v1). Continuing the trend we observe with other datasets, we find that \systemlong{} is 20.8$\times$ faster than \vanilla{} on CPU with no quality loss up to 100 passages. We do find that when $k=1000$ both \vanilla{} and \systemlong{} run out of memory on GPU; we believe we can address this in \system{} by implementing custom padding-free MaxSim kernels for GPU execution as discussed in \S\ref{subsection:plaid:padding_free_maxsim}. 

\subsection{Ablation} \label{subsection:eval:ablation}

\begin{figure}
    \centering
    \begin{subfigure}[b]{\columnwidth}
    \centering
    \includegraphics[scale=0.4]{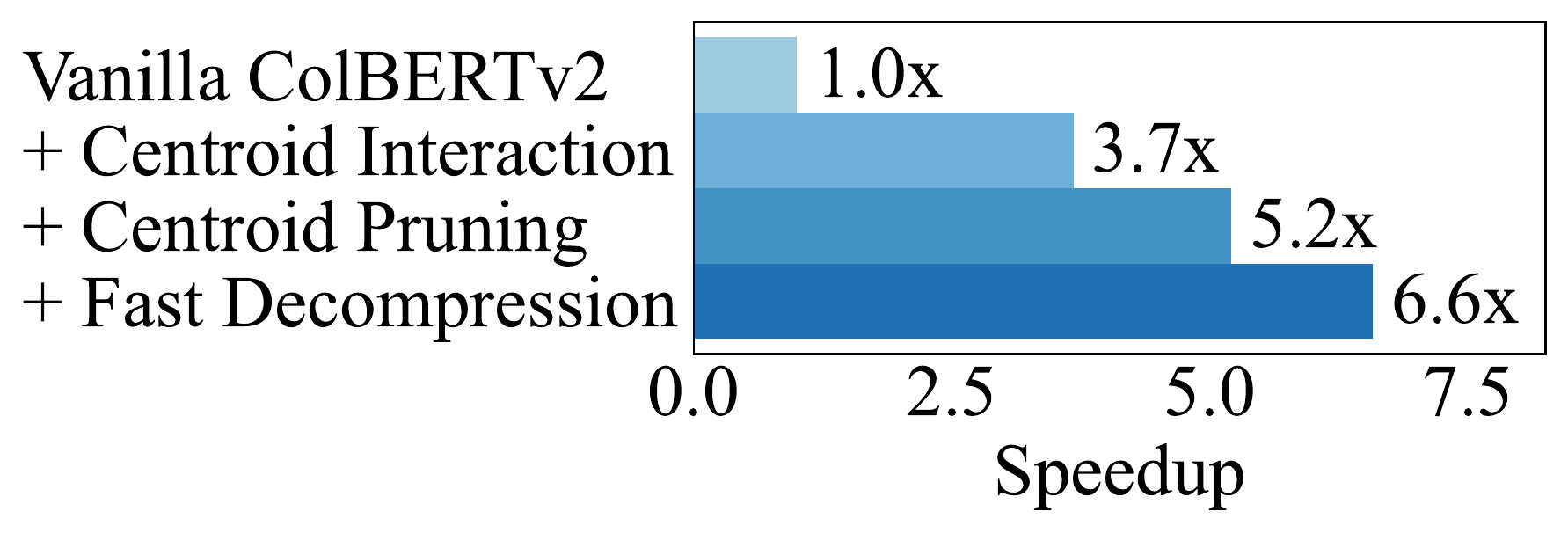}
    \caption{GPU.}
    \label{fig:gpu_ablation}
    \end{subfigure}
    \par\bigskip
    \par\bigskip
    \begin{subfigure}[b]{\columnwidth}
    \centering
    \includegraphics[scale=0.4]{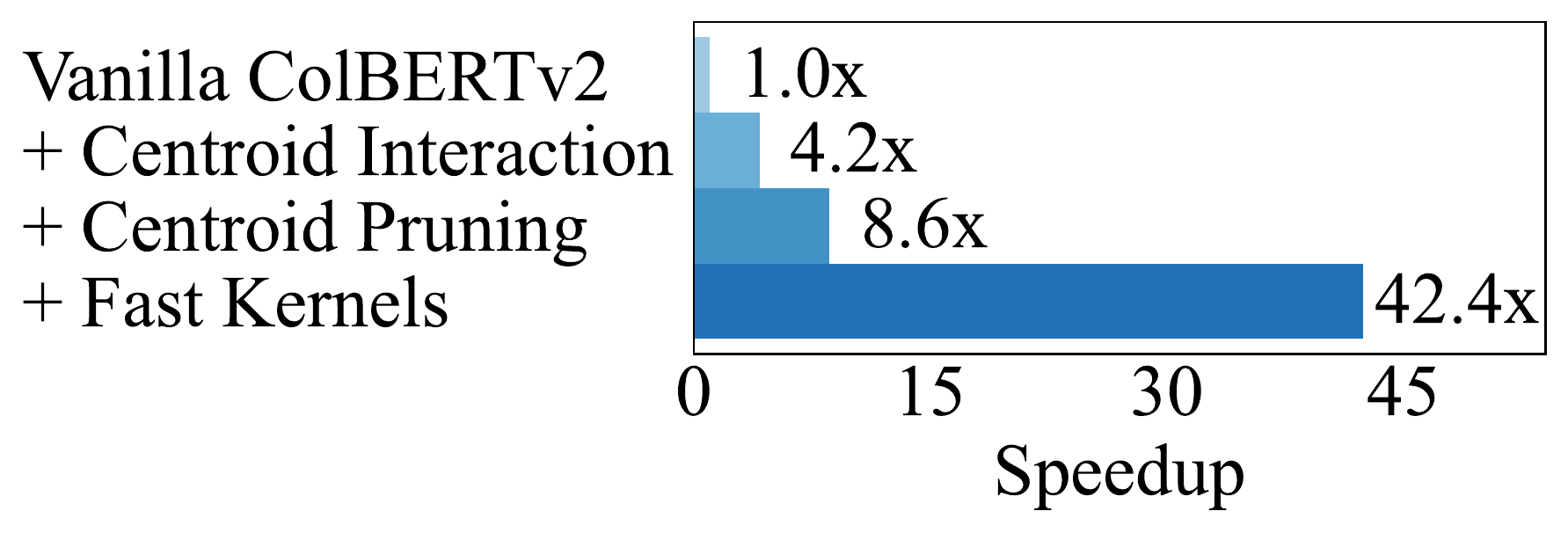}
    \caption{CPU (8 threads).}
    \label{fig:cpu_ablation}
    \end{subfigure}
    \caption{Ablation of performance optimizations included in \system{}.}
    \label{fig:ablation}
\end{figure}

Figure~\ref{fig:ablation} presents an ablation analysis to break down \system{}'s performance improvements for both GPU and CPU execution. Our measurements are taken from evaluation on a random sample of 500 MS MARCO v1 queries (note that this results in minor differences in the absolute numbers reported in Table~\ref{table:msmarco_end_to_end_results}). We consider \vanilla as a baseline, and then add one stage of centroid interaction without pruning (stage 3 in Figure~\ref{fig:scoring_pipeline}), followed by another stage of centroid interaction with centroid pruning (stage 2 in Figure~\ref{fig:scoring_pipeline}), and then finally the optimized kernels described in \secref{section:implementation}. When applicable we use hyperparameters corresponding to the $k=1000$ setting described in Table~\ref{table:plaid_parameters} (i.e., the most conservative setting).

We find that both the algorithmic improvements to the scoring pipeline as well as the implementation optimizations are key to \system{}'s performance. In particular, the centroid interaction stages alone deliver speedups of 5.2$\times$ on GPU and 8.6$\times$ on CPU, but adding the implementation optimizations result in additional speedups of 1.3$\times$ on GPU and 4.9$\times$ on CPU. Only enabling optimized C++ kernels on CPU without centroid interaction (not shown in Figure~\ref{fig:ablation}) results in an end-to-end speedup of just 3$\times$ compared to 42.4$\times$ with the complete \system{}.

\subsection{Scalability} \label{subsection:eval:scalability}

\begin{figure}
    \centering
    \includegraphics[scale=0.32]{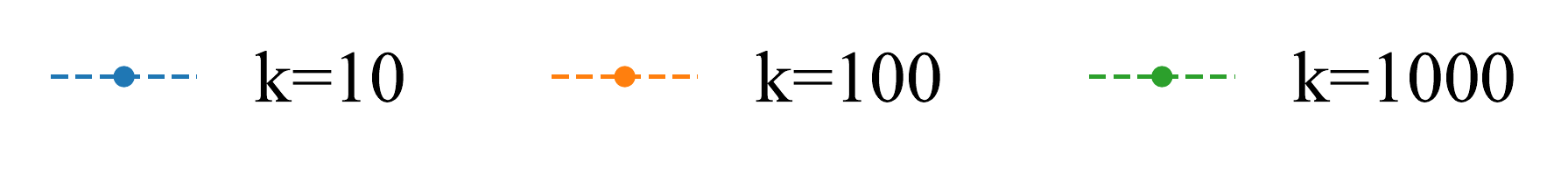}
    \begin{subfigure}[b]{0.48\columnwidth}
    \centering
    \includegraphics[width=1\columnwidth]{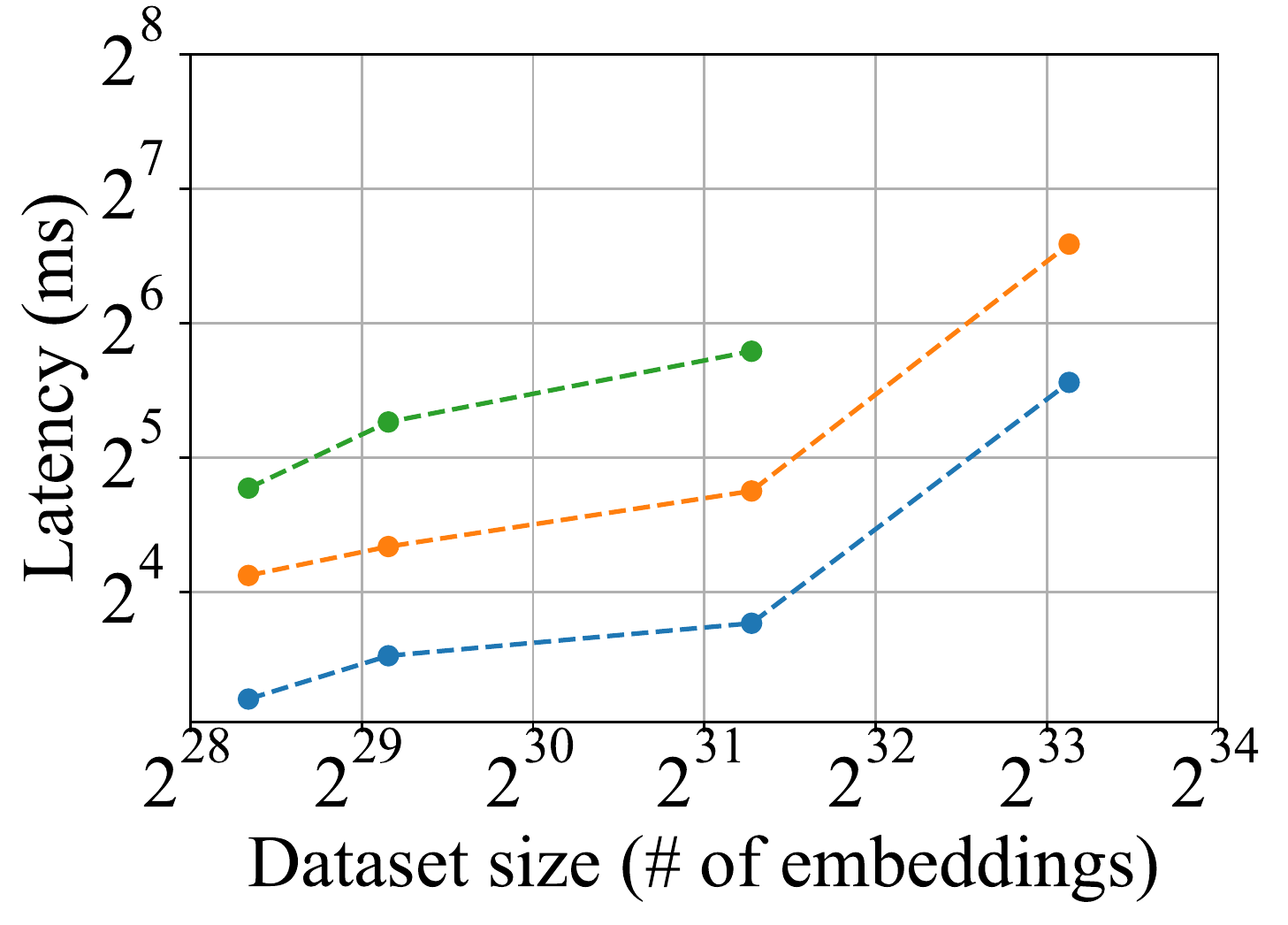}
    \caption{GPU.}
    \end{subfigure}
    \hfill
    \begin{subfigure}[b]{0.48\columnwidth}
    \centering
    \includegraphics[width=1\columnwidth]{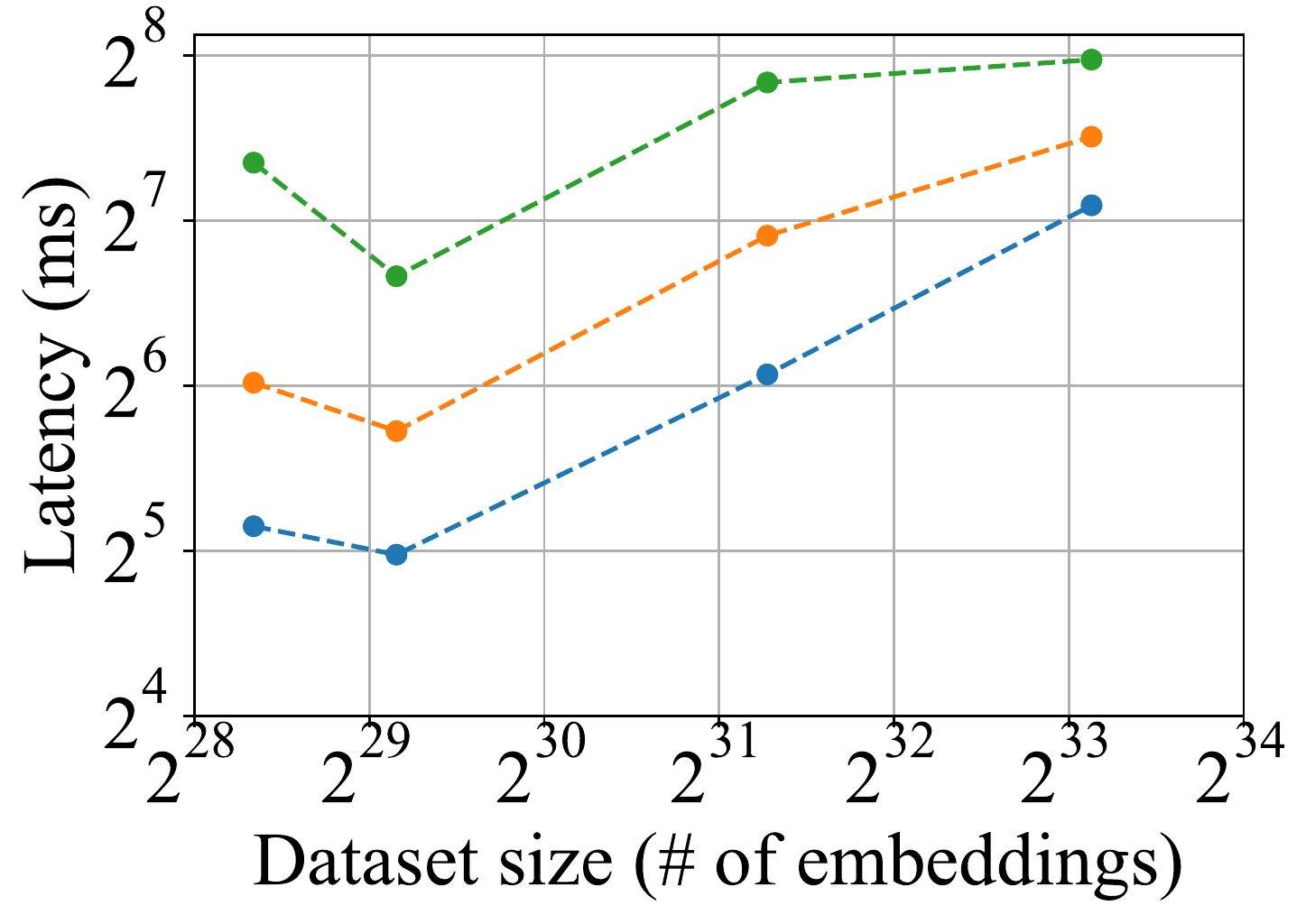}
    \caption{CPU (8 threads).}
    \end{subfigure}
    \caption{End-to-end latency versus dataset size (as measured in number of embeddings) for each setting of $k$ (note the log-log scale). Dataset sizes are taken from Table~\ref{table:eval_datasets}, and latency numbers are taken from Tables~\ref{table:msmarco_end_to_end_results},~\ref{table:wiki_end_to_end_results},~\ref{table:lotte_end_to_end_results}, and~\ref{table:msmarcov2_results}.} 
    \label{fig:latency_vs_dataset_size}
\end{figure}

\begin{figure}
    \centering
    \includegraphics[scale=0.4]{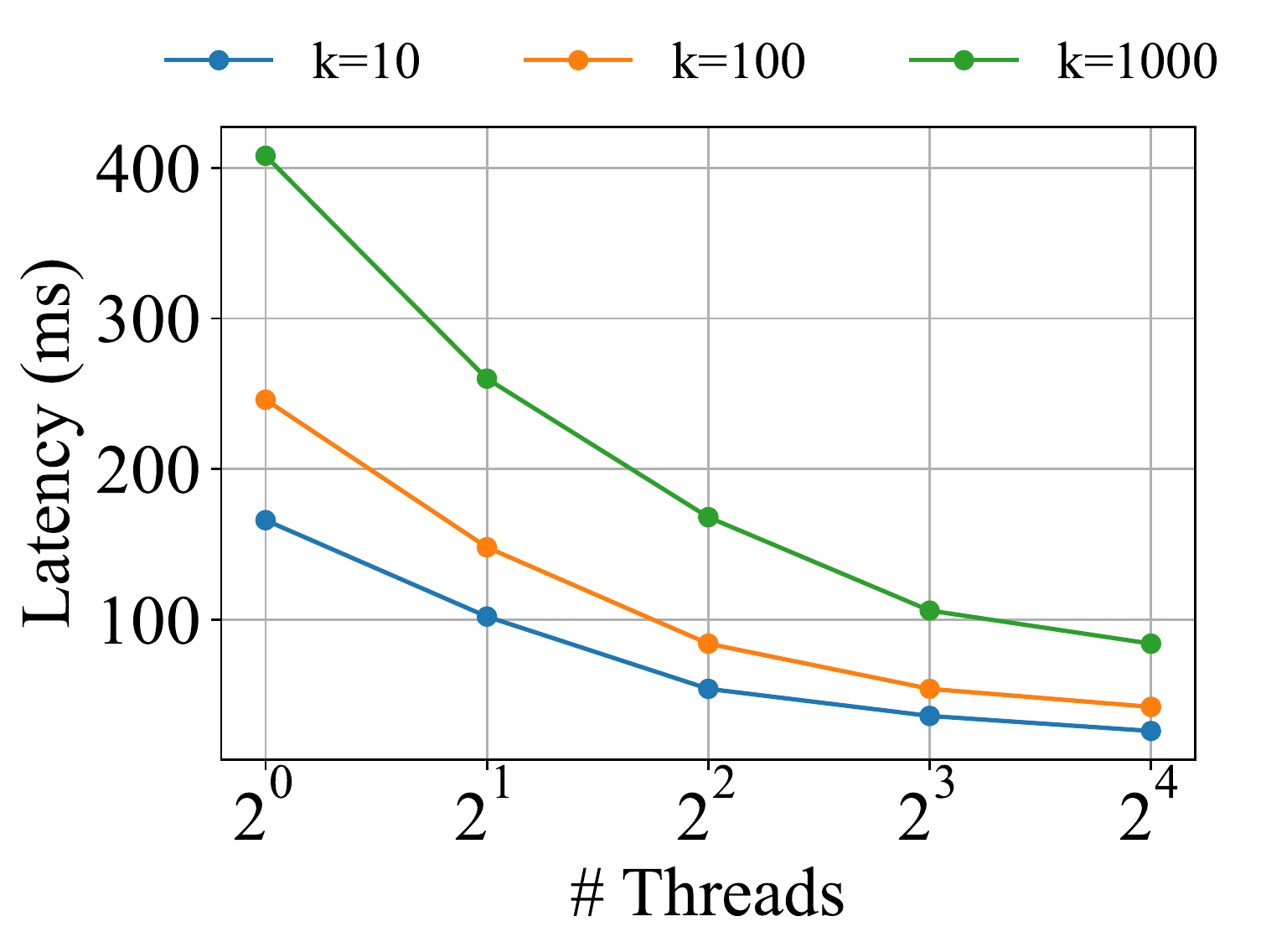}
    \caption{\system{} scaling behavior with respect to the number of available CPU threads.}
    \label{fig:cpu_scaling}
\end{figure}

We evaluate \system{}'s scalability with respect to both the dataset size as well as the parallelism degree (on CPU). 

First, Figure~\ref{fig:latency_vs_dataset_size} plots the end-to-end \systemlong{} latencies we measured for each benchmark dataset versus the size of each dataset (measured in number of embeddings). While latencies across different datasets are not necessarily directly comparable (e.g due to different passage lengths), we nevertheless aim to analyze high-level trends from this figure. We find that in general, \systemlong{} latencies appear to scale with respect to the square root of dataset size. This intuitively follows from the fact that ColBERTv2 sets the number of centroids proportionally to the square root of the number of embeddings, and the overhead of candidate generation is inversely correlated with the number of partitions. 

Next, Figure~\ref{fig:cpu_scaling} plots the latency achieved by \systemlong{} versus the number of available CPU threads, repeated for $k \in \{10, 100, 1000\}$. We evaluate a random sample of 500 MS MARCO v1 queries to obtain the latency measurements. We observe that \system{} is able to take advantage of additional threads; in particular, executing with 16 threads results in a speedup of 4.9$\times$ compared to single-threaded execution when $k=1000$. While \system{} does not achieve perfect linear scaling, we speculate that possible explanations could include remaining inefficiencies in the existing \vanilla{} candidate generation step (which we do not optimize at a low level for this work) or suboptimal load balancing between threads due to the non-uniform passage lengths. We defer more extensive profiling and potential solutions to future work.

\section{Conclusion} \label{section:conclustion}

In this work, we presented \system{}, an efficient engine for late interaction that accelerates retrieval by aggressively and cheaply filtering candidate passages. We showed that retrieval with only ColBERTv2 centroids retains high recall compared to \vanilla{}, and the distribution of centroid relevance scores skews toward lower magnitude scores. Using these insights, we introduced the technique of centroid interaction and incorporated centroid interaction into multiple stages of the \systemlong{} scoring pipeline. We also described our highly optimized implementation of \system{} that includes custom kernels for padding-free MaxSim and residual decompression operations. We found in our evaluation across several IR benchmarks that \systemlong{} provides speedups of 2.5--6.8$\times$ on GPU and 9.2--45$\times$ on CPU with virtually no quality loss compared to \vanilla{} while scaling effectively to a dataset of 140 million passages.

\bibliographystyle{ACM-Reference-Format}
\bibliography{paper}

\end{document}